\documentclass[conference]{IEEEtran} 
\pdfoutput=1

\IEEEoverridecommandlockouts

\usepackage{graphicx,cite}
\usepackage{amsmath}
\usepackage{url}

\usepackage{amssymb}
\usepackage{amsfonts}
\usepackage{lineno}
\usepackage{dsfont}
\usepackage{bbm}

\usepackage{soul}

\usepackage{times}
\usepackage{algorithmic}
\usepackage{algorithm}

\newtheorem{definition}{\bf Definition}

\DeclareMathOperator*{\argmax}{arg\,max}
\newcommand{\vect}{\boldsymbol}
\newcommand{\vecta}[2]{#1_1,#1_2,\ldots,#1_{#2}}
\newcommand{\vectab}[2]{#1_1,\ldots,#1_{#2}}
\newcommand{\Seta}[1]{1,\ldots,#1}
\newcommand{\Set}[1]{\mathcal{#1}}

\newcommand{\texth}[1]{\!^\text{#1}}

\newcommand{\vectx}{\vect{x}}

\begin{document}

\title{\huge Opportunistic Sleep Mode Strategies  in \\ Wireless Small Cell Networks
}

\author{
\IEEEauthorblockN{Sumudu Samarakoon\IEEEauthorrefmark{1}, Mehdi Bennis\IEEEauthorrefmark{1}, Walid Saad\IEEEauthorrefmark{2} and Matti Latva-aho\IEEEauthorrefmark{1} \\}
\IEEEauthorblockA{\small\IEEEauthorrefmark{1}Centre for Wireless Communications, University of Oulu, Finland, \\ email: \{sumudu,bennis,matti.latva-aho\}@ee.oulu.fi \\
\IEEEauthorrefmark{2}Electrical and Computer Engineering Department, University of Miami, Coral Gables, FL, USA, email: walid@miami.edu}
\vspace{-25pt}
\thanks{This work is supported by the SHARING project under the Finland grant 128010 and the U.S. National Science Foundation (NSF) under Grant CNS-1406947 and CNS-1253731.}
}

\maketitle
\nopagebreak[4]
\begin{abstract}
The design of energy-efficient mechanisms is one of the key challenges in emerging wireless small cell networks. 
In this paper, a novel approach for opportunistically switching ON/OFF base stations to improve the energy efficiency in wireless small cell networks is proposed. 
The proposed approach enables the small cell base stations to optimize their downlink performance while balancing the load among each another, while satisfying their users' quality-of-service requirements. 
The problem is formulated as a noncooperative game among the base stations that seek to minimize a cost function which captures the tradeoff between energy expenditure and load.
To solve this game, a distributed learning algorithm is proposed using which the base stations autonomously choose their optimal transmission
strategies. 
Simulation results show that the proposed approach yields significant performance gains in terms of reduced energy expenditures up to $23\%$ and reduced load up to $40\%$ compared to conventional approaches.

\end{abstract}

\section{Introduction}\label{sec:introduction}

In the past decade, the demand for wireless resources has grown exponentially due to the proliferation of bandwidth-intensive applications such as video streaming and social media~\cite{online:comscore13}.
This growth increases the load on existing wireless cellular systems and leads to an increased energy consumption over the radio spectrum~\cite{pap:arshad12,pap:brevis11}. 
Therefore, developing energy-efficient mechanisms for resource allocation in wireless networks has become a major research topic in recent years~\cite{pap:brevis11,jnl:kyuho11}.
In this respect, one promising approach to provide high wireless quality-of-service~(QoS) while maintaining energy-efficient operation is through the deployment of low-cost, low-power small cells over existing cellular networks.

Existing literature has studied a number of problems related to resource allocation in small cell networks, such as base station~(BS) placement, load balancing, power control, and dynamic BS sleep-wake mechanism, among others \cite{pap:arshad12,pap:brevis11,pap:richter12,jnl:hongseok12,jnl:kyuho11,pap:zhou11}.
In \cite{pap:arshad12}, an optimal deployment strategy is proposed for a two-tier network based on power consumption minimization subject to a target spectral efficiency.
In \cite{pap:brevis11}, a stochastic programing approach with the goal of minimizing energy consumption is proposed for optimizing micro-BS locations.
The authors in \cite{pap:richter12} propose a decentralized cellular deployment mechanism based on forced fields for load balancing.  
A distributed load balancing problem is studied in~\cite{jnl:hongseok12} by introducing a utility function which captures different user association policies.
In \cite{jnl:kyuho11}, a BS operation mechanism is proposed based on tradeoffs between energy and traffic load.
The proposed greedy algorithm improves the energy efficiency by allowing certain BSs to switch between on and off states.
A probabilistic approach for sleep-wake mechanism is presented in~\cite{pap:zhou11} to optimize the energy efficiency of relays in conventional cellular networks.

Although the above studies provide good insights on improving energy efficiency, all of them rely on a central controller which gathers all network information and makes all decisions.
Such a mechanism introduces additional costs and overhead for information exchange and backhauls.
Therefore, providing self-organizing capabilities to BSs seems to be a cost effective solution.
However, with the absence of a central controller, the main issue faced by the BS activation/deactivating mechanism is to determine when and how to activate the sleeping BSs.
This requires monitoring the entire network, the capability to predict the load, and react to the network changes.
Failure to activate a sufficient number of BSs causes outages and activation of excessive number of BSs degrades the network efficiency.
In this paper, our goal is to overcome these challenges by developing a new approach for opportunistic sleep mode strategies in small cell networks.

The main contribution of this paper is to develop opportunistic on/off strategies for BSs allowing them to decide on whether to switch to a sleep mode or to an active mode, depending on the current traffic load and network environment.
Within the context of small cell networks, developing such a dynamic state switching algorithm requires a self-organizing, decentralized approach so as to  minimize the overhead or coordination among base stations.
Unlike the previous studies~\cite{jnl:kyuho11,pap:zhou11} which rely on a central controller, we propose a novel implicit coordination mechanism for dynamically activating/deactivating  base stations while maintaining a balance between throughput and energy consumption.
In the proposed approach, the BSs learn their best strategy profile based only on their individual energy consumption and handled load, without requiring global information. 
The goal of the BSs is to minimize a cost function which captures both energy consumption and load. 
We cast the problem as a non-cooperative game between the BSs.
To solve this game, we propose a distributed algorithm using notions from regret learning~\cite{pap:bennis12,jnl:qian12}.
Simulation results show that the proposed approach improves the energy efficiency and reduces the overall load in the system as compared to conventional approaches.

The rest of the paper is organized as follows. The system model and problem formulation are presented in Section~\ref{sec:system_model_problem_form}.  
The proposed game theoretical approach is discussed and the decentralized ON/OFF algorithm is proposed in Section~\ref{sec:solution}. 
Simulation results are presented and analyzed in Section~\ref{sec:results}. 
Finally, conclusions are drawn in Section~\ref{sec:conclusions}.
\section{System Model and Problem Formulation}\label{sec:system_model_problem_form}

\subsection{Notation}\label{subsec:notation}

The regular symbols represent the scalars while the boldface symbols are used for vectors.
The sets are denoted by upper case calligraphic symbols.
$|\Set{X}|$ and $\Delta(\Set{X})$ represent the cardinality and the set of all probability distributions of the finite set $\Set{X}$, respectively.
The function $\mathbbm{1}_{\Set{X}}(x)$ denotes the indicator function which is defined as,
$$\mathbbm{1}_{\Set{X}}(x)=\begin{cases} 1 &\mbox{if } x\in\Set{X}, \\ 0 &\mbox{if } x\notin\Set{X}.\end{cases}$$

\subsection{Network Model}\label{subsec:network_model}

\begin{figure}[!t]
\centering
\includegraphics[width=.44\textwidth]{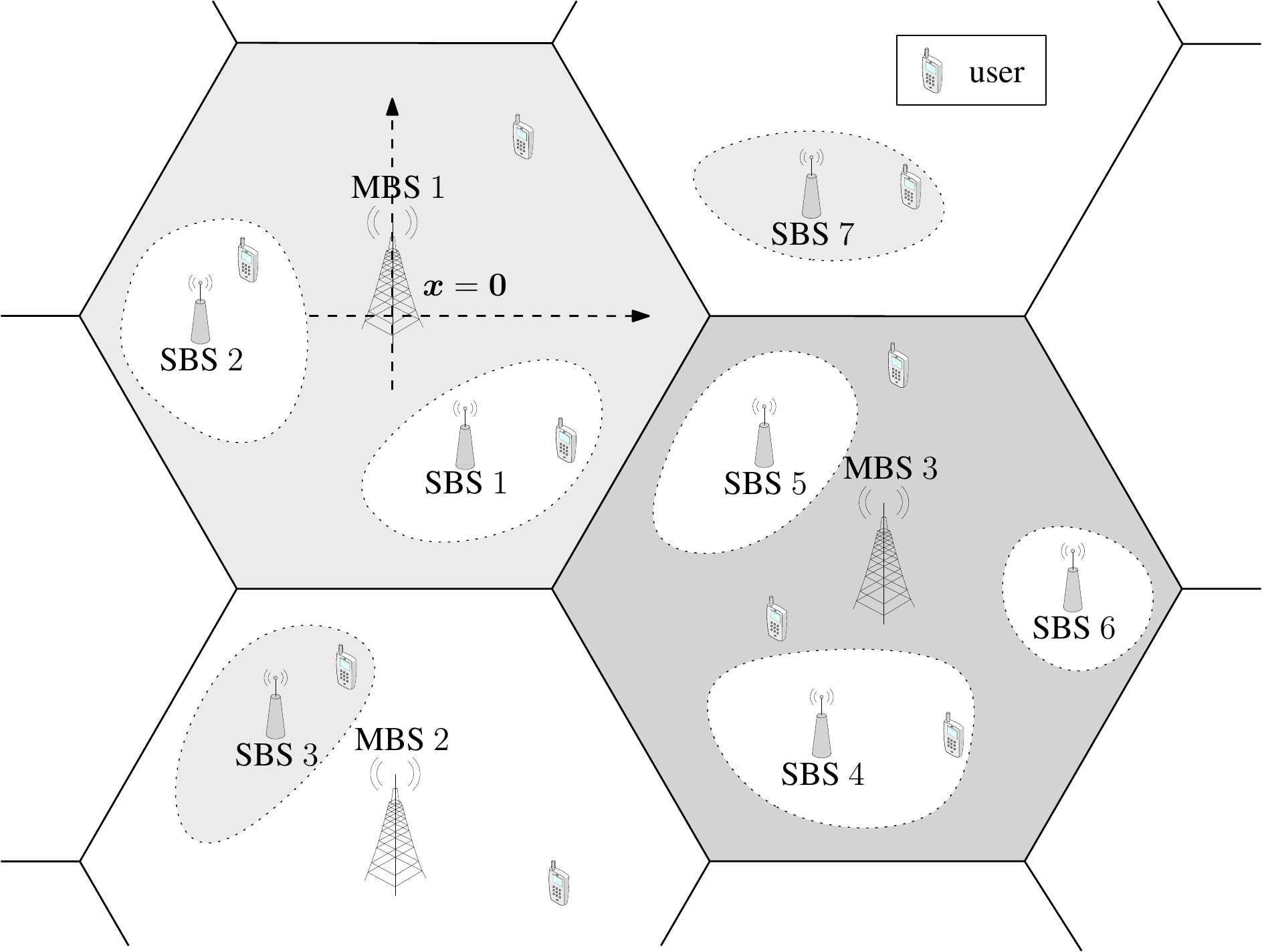}
\caption{Illustration of the studied network. MBS-$1$ located at the origin and the shaded regions around the BSs represent their respective coverage areas.  }
\label{fig:network_layout}
\end{figure}

Consider the downlink transmission of a heterogeneous wireless network with a set of BSs $\Set{B}=\{\Seta{B}\}$.
The set $\Set{B}$ consists of a set of small cell base stations (SBSs) $\Set{B}_S=\{\Seta{B_S}\}$ underlaid on a macro cellular network with a set of macro base stations (MBSs) $\Set{B}_M=\{\Seta{B_M}\}$, i.e. $\Set{B}=\Set{B}_M\cup\Set{B}_S$.
Without loss of generality, we assume that the MBS $1$ is located at the origin of the two-dimensional network layout and we let $\vectx$ be any location on the plane measured with respect to the origin.
Moreover, let $\Set{L}_b$ be the coverage area of BS $b$ such that any given user equipment (UE) at a given location $\vectx$ is served by BS $b$ if $\vectx\in\Set{L}_b$.
An illustrative example is shown in Fig~\ref{fig:network_layout}.

Let $S_b$ be the transmission indicator of BS $b$ such that $S_b=1$ indicates the active state while $S_b=0$ reflects the idle or sleep state.
Here, we assume that, in active state, each BS will service all UEs in its coverage area.
From an energy saving perspective, some BSs might have an incentive to switch into sleep mode.
Note that during the idle state, a BS consumes nonzero power to sense the UEs in its vicinity. 
The power consumption of BS $b$ is given by~\cite{pap:georgios12B}:
\begin{equation}\label{eqn:state_based_power}
	P^{\texth{Total}}_b =
	\begin{cases}
		P_b^{\texth{Idle}} = \frac{ P_{\text{RF}}+P_{\text{BB}} }{ (1-\sigma_{\text{DC}}) (1-\sigma_{\text{MS}}) (1-\sigma_{\text{cool}}) } = \frac{P_{\text{RF}}+P_{\text{BB}}}{\sigma} &\mbox{if}~S_b=0, \\
 		\big( P_b^{\texth{Work}} + P_b^{\texth{Idle}} \big) =  \frac{P_b}{\eta\sigma(1-\sigma_{\text{feed}})}+ P_b^{\texth{BCK}} + P_b^{\texth{Idle}}&\mbox{if}~S_b=1,
	\end{cases}
\end{equation}
where $P_{\text{RF}},~P_{\text{BB}}$ represent, respectively, the total power of the radio frequency component and baseband unit, $\sigma_{\text{feed}},~\sigma_{\text{DC}},~\sigma_{\text{MS}},~\sigma_{\text{cool}}$ are respective loss fractions in feeders, DC-DC conversion, main supply and cooling, and $\eta$ is the efficiency of the power amplifier.
$P_b^{\texth{BCK}}$ represents the power consumption for the backhaul of $b\in\Set{B}$.
We note that, with this model, we are able to capture not only the power consumption due to the transmission, but also the power consumption required to maintain the BS in active mode.

Consider that BS $b$ uses a cell range expansion bias (CREB) $\zeta_b$ to absorb additional UEs (expand its coverage $\Set{L}_b$) along the transmission power $P_b=P_b^{\texth{Work}}$.
The concept of CREB has been proposed in small cell networks due to the disparate cell sizes between different BSs~\cite{book:quek13}.
Since the CREB concept is used only by SBSs, we let $\zeta_b=0$ for all $b\in\Set{B}_M$. 
Moreover, we assume that all BSs transmit on the same frequency spectrum (i.e., co-channel deployment).
Therefore, the received signal to interference and noise ratio (SINR) from BS $b$ at location $\vectx\in\Set{L}_b$ is given by:
\begin{equation}\label{eqn:sinr}
	\gamma_b(\vectx) = \frac{ P_{b}S_{b}h_{b}(\vectx) }{ \sum_{\forall b'\in\Set{B}/b} P_{b'}S_{b'}h_{b'}(\vectx) + N_0 },
\end{equation}
where $h_{b'}(\vectx)$ is the channel gain from BS $b'$ to a given UE location at $\vectx$ and $N_0$ is the noise variance.
Further, the data rate at a given location $\vectx$ from BS $b$ is given by:
\begin{equation}\label{eqn:rate_ue}
	R_b(\vectx) = \omega\log_2\big(1 + \gamma_b(\vectx)\big),
\end{equation}
where $\omega$ is the bandwidth.

We assume that the UEs connected to BS $b$ are heterogeneous in nature such that each UE has a different QoS requirement based on its individual packet arrival rate.
In this respect, let $\lambda(\vectx)$ and $1/\mu(\vectx)$ be the packet arrival rate and the mean packet size of any UE at location $\vectx\in\Set{L}_b$.
The data rate offered to the UE at location $\vectx$ from BS $b$ is $R_b(\vectx)$ and thus, the load density of BS $b$ becomes $\vect{\varrho}_b = \{\varrho_b(\vectx)|\vectx\in\mathcal{L}_b\}$ where $\varrho_b(\vectx)=\frac{\lambda_b(\vectx)}{\mu_b(\vectx)R_b(\vectx)}$.
Consequently, the BS load $\rho_b$ of a BS $b$ is given by:
\begin{equation}\label{eqn:load_bs}
	\rho_b = \int_{\vectx\in\Set{L}_b} \varrho_b(\vectx) d\vectx. 
\end{equation}

\subsection{Problem Formulation}\label{sec:formulations}

The configuration of the entire network is defined by the transmission powers and the states of all the BSs.
This configuration can thus be captured by a transmission power vector $\vect{P}$, a CREB vector $\vect{\zeta}$, and a state indicator vector $\vect{S}$ as follows;
\begin{equation} \label{eqn:parameter_vectors}
	\vect{P} = [\vectab{P}{B}],
	~\vect{\zeta} = [\vectab{\zeta}{B_s}],
	~\vect{S} = [\vectab{S}{B}].
\end{equation}
Due to the fact that the load $\rho_b$ of each BS $b\in\Set{B}$ is coupled with the above parameters as per (\ref{eqn:state_based_power})-(\ref{eqn:load_bs}), we map the configuration $(\vect{P},\vect{\zeta},\vect{S})$ to a vector $\vect{\rho}=[\vecta{\rho}{B}]$.
Hereinafter, we refer to the load $\vect{\rho}$ as the ``network configuration".

For a given network configuration $\vect{\rho}$, BS $b$ handles the load $\rho_b$ from the set of UEs in its coverage area $\Set{L}_b$.
Consider a scenario in which BS $b$ increases its transmission power from $P_b$ to $Q_b$ in order to serve the UEs in $\Set{L}_b$.
Following (\ref{eqn:sinr}) and (\ref{eqn:rate_ue}), this power increment results in a higher SINR $\gamma_b(\vectx,Q_b)>\gamma_b(\vectx,P_b)$ and higher data rate $R_b(\vectx,Q_b)>R_b(\vectx,P_b)$ at location $\vectx\in\Set{L}_b$.
Furthermore, for all $\vectx\in\Set{L}_b$, the load density $\varrho_b(\vectx,Q_b)$ of  BS $b$ decreases resulting in a load reduction,
$$\rho_b(Q_b) = \int_{\vectx\in\Set{L}_b} \varrho_b(\vectx,Q_b) d\vectx < \int_{\vectx\in\Set{L}_b} \varrho_b(\vectx,P_b) d\vectx = \rho_b(P_b).$$
Clearly, for a given coverage area $\Set{L}_b$, each BS $b$ can reduce the handled load $\rho_b$ by increasing the offered throughput $R_b(\vectx)$ for all $x\in\Set{L}_b$.
However, in order to achieve higher throughputs, the BS needs to increase its power consumption.
Thus, there is a tradeoff between  load reduction (throughput increment and delay reduction) and  energy consumption reduction.

Here, for each BS $b\in\Set{B}$, we define a cost function that captures both energy consumption and load, as follows:
\begin{equation}\label{eqn:bs_utility_metric}
	\Gamma_b(\vect{\rho}) = \alpha_b \underbrace{P_b^{\texth{Total}}(\vect{\rho})}_{\text{energy}} \; + \; \beta_b \underbrace{\rho_b}_{\text{load}},
\end{equation}
where the coefficients $\alpha_b$ and $\beta_b$ are weight parameters that indicate the impact of energy and load on the cost, respectively.
The overall objective is to minimize the total network cost:
\begin{eqnarray}\label{eqn:optimization_energy_efficiency}
	 \underset{\vect{\rho}}{\text{minimize}} && \sum_{\forall b\in\mathcal{B}} \Gamma_b(\vect{\rho}) \\
	 \label{cns:load} \text{subject to} &&
			0 \leq \rho_b \leq 1, \quad \forall b\in\Set{B} \\
			\label{cns:power} && P_b^{\texth{Total}}(\vect{\rho}) \leq P_b^{\texth{Max}}, \quad \forall b\in\Set{B} 
\end{eqnarray}
where $P_b^{\texth{Max}}$ is the maximum transmit power of BS $b$.
The load constraint (\ref{cns:load}) avoids outages and ensures service for all UEs in $\Set{L}_b$ serviced by BS $b$.

\section{Self-Organizing Switching ON/OFF Mechanism}\label{sec:solution}

The goal of this work is to propose a self-organizing solution for (\ref{eqn:optimization_energy_efficiency})-(\ref{cns:power}) in which each BS individually adjust its transmission parameters $\vect{\rho}_b$, without global network informations.
We assume that BSs do not communicate among each others, 
thus, each BS makes its decision independently.
To do so, we use a regret-based learning approach~\cite{pap:bennis12,jnl:qian12}, in which the proposed solution consists of two interrelated parts: user association and BS transmission optimization.

\subsection{Load-Based User Association}

The coverage area $\Set{L}_b$ defines the locations of UEs associated with BS $b$. 
Classical UE association techniques include received signal strength indication (RSSI) and SINR based associations~\cite{pap:richter12}.
In these two cases, a UE connects to the BS which offers the best signal strength (maximum RSSI based approach) or to the BS providing the best SINR (maximum SINR based approach).
However, these two techniques are oblivious to the base stations' and the network's traffic load. 
This may lead to overloading BSs and lower spectral efficiencies.
Thus, a smarter mechanism in which the BSs advertise their load to all UEs within their coverage area is desirable.
Such association mechanism allows UEs to consider both BS capability (load) and the quality of the communication link.
Moreover, with such approach, the system yields fewer number of unsatisfied UEs and lesser overloaded BSs.

At time instant $t$, each BS $b$ advertises its estimated load $\hat{\rho}_b(t)$ and an offset $\epsilon_b$ via a broadcast control message along with its transmission power $P_b(t)$ and CREB $\zeta_b(t)$.
Considering both the received signal strength and load, at time $t$ the UE at location $\vectx$ connects to BS $b(\vectx,t)$, $\vectx\in\Set{L}_{b(\vectx, t)}$, according to the following UE association rule:
\begin{equation}\label{eqn:ue_association}
	b(\vectx, t) = \argmax_{b \in B} \Big\{ \Big(\hat{\rho}_b(t)+\epsilon_b\Big)^{-\delta}P_b^{\texth{Rx}}(t) \Big\}.
\end{equation}
Here, $P_b^{\texth{Rx}}(t) = \big( P_b(t) + \mathbbm{1}_{\Set{B}_s}(b) \zeta_b(t) \big)S_b(t) h_{b}(\vectx,t)$ is the received signal power at the UE in location $\vectx$ from BS $b$ at time $t$.
The impact of the load is determined by the coefficient $\delta\geq 0$.
The classical RSSI-based UE association is a special case of (11) when $\delta = 0$.
Note that the offset $\epsilon_b$ is selected as $\epsilon_b = 1 - \rho^{\texth{Best}}_b$ where $\rho^{\texth{Best}}_b$ is the preferred load per BS $b$.
The UE association rule (\ref{eqn:ue_association}) encourages UEs to connect to underloaded BSs $\big(\hat{\rho}_b(t)\leq\rho^{\texth{Best}}_b\big)$.

Due to fact that the BSs need to estimate their loads beforehand, the estimations must accurately reflect the actual load.
In order to obtain an accurate estimation for the load of the BS $b$, we compute the load estimation $\hat{\rho}_b(t)$ at time $t$ based on history as follows:
\begin{equation}\label{eqn:load_estimation}
	\hat{\rho}_b(t) = \hat{\rho}_b(t-1) + \nu(t) \Big( \rho_b(t-1)-\hat{\rho}_b(t-1) \Big),
\end{equation}
where $\nu(t)$ is the learning rate of the load estimation.
Leveraging different time-scales, we assume that $\nu(t)$ is selected such that the load estimation procedure~(\ref{eqn:load_estimation}) is much slower than the UE association process.
In addition, this learning rule considers both the instantaneous load $\big(\rho_b(t-1)\big)$ and the long-term history $\big(\hat{\rho}_b(t-1)\big)$, and to predict the load by balancing between those two quantities.

\subsection{Game Formulation}

In the proposed approach, the BSs need to autonomously select $\vect{\rho}$ in order to minimize their cost functions.
However, the cell coverage and the achievable throughput of a BS depends not only on its own choice of action but also on remaining BSs due to the interference.
In this regard, we formulate a non-cooperative game $\Set{G} = \big( \Set{B}, \{\Set{A}_b\}_{b\in\Set{B}}, \{{u}_b\}_{b\in\Set{B}} \big)$ in which the set of BSs ($\Set{B}$) is the set of players.
Each player $b\in\Set{B}$ has $\Set{A}_b=\big\{ a_{b,1},\ldots,a_{b,|\Set{A}_b|} \big\}$ set of actions where an action of BS $b$, $a_b$, is composed of its transmission power $P_b\in[0,P_b^{\texth{Max}}]$, CREB $\zeta_b\in[0,\zeta_b^{\texth{Max}}]$, and state $S_b\in\{0,1\}$, i.e. $a_b=(P_b,\zeta_b,S_b)$.
The action $a_b$ of BS $b$ and the actions of the other BSs $\vect{a}_{-b}$ define the load $\vect{\rho}$ of the system.
$u_b$ is the utility function of BS $b$ with $u_b:\Set{A}_b\rightarrow\mathbb{R}$ where $u_b(a_b,\vect{a}_{-b}) = -\Gamma_b(\vect{\rho})$.
Let $\vect{\pi}_b(t) = \big[ \pi_{b,1}(t),\ldots,\pi_{b,|\Set{A}_b|}(t) \big]$ be a probability distribution in which BS $b$ selects a given action from $\Set{A}_b$ at time instant $t$, i.e. $\pi_{b,i}(t)=\mbox{Pr}\big(a_b(t)=a_{b,i}\big)$ is BS $b$'s \emph{mixed strategy} where $a_b(t)$ is the action of player $b$ at time $t$.
Our goal is to develop a distributed mechanism to solve the switch ON/OFF game and reach the $\varepsilon$-coarse correlated equilibrium ($\varepsilon$-CCE) defined as follows~\cite{jnl:qian12}:
\begin{definition}
	({\it $\varepsilon$-coarse correlated equilibrium}):
	A mixed strategy probability $\vect{\pi}_b$ is  an $\varepsilon$-coarse correlated equilibrium if, $\forall b\in\Set{B}$ and $\forall a_b'\in\mathcal{A}_b$, it holds that:
\small
\begin{multline*}
	\sum_{\vect{a}_{-b}\in\mathcal{A}_{-b}} \Biggl( u_b{(a_b',\vect{a}_{-b})} {\pi}_{-b,\vect{a}_{-b}} \Biggr)  
	- \sum_{\vect{a}\in\vect{\mathcal{A}}} \Biggl( u_b{(\vect{a})}{\pi}_{b} \Biggr) \leq \varepsilon,
\end{multline*}
\normalsize
	where ${\pi}_{-b,\vect{a}_{-b}}=\sum_{\forall a_b\in\mathcal{A}_b}\pi(a_b,\vect{a}_{-b})$ is the marginal probability distribution w.r.t. $a_b$ and $\vect{a}=[\vectab{a}{B}]$.
\end{definition}

In order to reach the $\varepsilon$-CCE, first, suppose that a given BS $b$ constantly changes its actions following a particular strategy $\vect{\pi}_b$ and observes the time-average of its utility $\overline{u}_b(t)$ while the rest of the players follow their strategies captured by vector $\vect{\pi}_{-b}$.
While BS $b$ plays action $a_b(t)$, it may regret or be satisfied about the action it played based on the observed utility feedback $\overline{u}_b(t)$. 
Therefore, player $b$ estimates its utility ${\vect{\hat{u}}}_b(t)=\big[ \hat{u}_{b,1}(t),\ldots,\hat{u}_{b,|\Set{A}_b|}(t) \big]$ and regret ${\vect{\hat{r}}}_b(t)=\big[ \hat{r}_{b,1}(t),\ldots,\hat{r}_{b,|\Set{A}_b|}(t) \big]$ for each action assuming it has played the same action during all previous times $\{\Seta{t-1}\}$.
At each time $t$, player $b$ updates its mixed strategy probability distribution $\vect{\pi}_b$ in which the actions with higher regrets are exploited while exploring the actions with low regrets~\cite{pap:bennis12}.
Such behavior is captured by the Boltzmann-Gibbs (BG) distribution $\big(\vect{G}_b=[G_{b,1},\ldots,G_{b,|\Set{A}_b|})\big]$ which is calculated as follows:
\begin{equation}
	G_{b,i}\big(\vect{\hat{r}}_b(t)\big) = \frac{\exp\big(\kappa_b \hat{r}_{b,i}(t)\big)} {\sum_{\forall i'\in\Set{A}_b} \exp\big(\kappa_b \hat{r}_{b,i'}(t)\big) }, \: i\in\Set{A}_b
\end{equation}
where $\kappa_b>0$ is a temperature parameter which balances between exploration and exploitation.
For each time $t$, all the estimations for any player $b\in\Set{B}$, $\vect{\hat{u}}_b(t),~\vect{\hat{r}}_b(t)$ and $\vect{\pi}_b(t)$, are updated as follows;
\begin{equation}\label{eqn:algoUpdates}
\begin{cases}
	{\hat{u}}_{b,i}(t) &= {\hat{u}}_{b,i}(t-1)  \\
	&\hfill + \tau_b(t-1)\mathds{1}_{\{a_{b,i}=\rho_b(t-1)\}} \Bigl({u}_b(t-1)-{\hat{u}}_{b,i}(t-1)\Bigr),\\
	{\hat{r}}_{b,i}(t) &= {\hat{r}}_{b,i}(t-1)  \\
	& +\iota_b(t) \Bigl({\hat{u}}_{b,i}(t-1)-{u}_b(t-1)-{\hat{r}}_{b,i}(t-1)\Bigr),\\
	\pi_{b,i}(t) &= \pi_{b,i}(t-1) \\
	& + \varepsilon_b(t) \Bigl(G_{b,i}\big({\vect{\hat{r}}}_b(t-1)\big)-\pi_{b,i}(t-1)\Bigr).
\end{cases}
\end{equation}
with the learning rates $\tau,~\iota$ and $\varepsilon$ satisfying,
\small
\begin{multline}\label{eqn:learningRateConditions}
	(i) \quad \displaystyle\lim_{t\rightarrow\infty}\sum_{n=1}^t\tau(n) = +\infty, \quad   \displaystyle\lim_{t\rightarrow\infty}\sum_{n=1}^t\iota(n) = +\infty \\
	 \mbox{and} \quad \displaystyle\lim_{t\rightarrow\infty}\sum_{n=1}^t\varepsilon(n) = +\infty.
\end{multline}
\vspace{-7 mm}
\begin{multline}
	(ii) \quad \displaystyle\lim_{t\rightarrow\infty}\sum_{n=1}^t\tau^2(n) < +\infty, \quad \displaystyle\lim_{t\rightarrow\infty}\sum_{n=1}^t\iota^2(n) < +\infty, \\
	\mbox{and} \quad \displaystyle\lim_{t\rightarrow\infty}\sum_{n=1}^t\varepsilon^2(n) < +\infty.
\end{multline}
\vspace{-7 mm}
\begin{multline}
(iii) \quad \displaystyle\lim_{t\rightarrow\infty}\frac{\iota(t)}{\tau(t)} = 0 \quad \mbox{and} \quad \displaystyle\lim_{t\rightarrow\infty}\frac{\varepsilon(t)}{\iota(t)} = 0,
\end{multline}
\normalsize
This process guarantees the convergence of the algorithm to an $\epsilon$-CCE~\cite{pap:bennis12,jnl:qian12}.
Our choice of the learning rates follows the format of ${1}/{t^c}$ with exponent $c$.

At the beginning of each time instant $t$, all BSs advertise their loads $(\hat{\rho}_b)$ and select their actions based on their mixed-strategy probabilities, i.e. for all $b\in\Set{B}$, 
\begin{equation}\label{eqn:prob_to_action}
	a_b(t) = f\big( \pi_b(t-1) \big)
\end{equation}
where $f:\pi_b\rightarrow a_{b,i}$ is the mapping from probability distribution to an action.
Based on the the estimated load, UEs associate as per $a_b(t)$.
All the BSs carry out the transmission based on their actions $\vect{a(t)}$ and calculate their utilities $u_b(t) = -\Gamma_b(t)$.
Each BS individually updates its utility, regret, and load estimations $\big(\vect{\hat{u}}_b(t),\vect{\hat{r}}_b(t),\vect{\hat{\rho}}_b(t)\big)$ along with the mixed strategy probabilities $\big(\pi_b(t)\big)$.
The proposed algorithm is summarized in Algorithm \ref{alg:active_deactive}.

\begin{algorithm}[!t]
\caption{Opportunistic Switching ON/OFF Algorithm}
\label{alg:active_deactive}  
\begin{algorithmic}[1]                    
	\STATE {\bf Input:} $\vect{\hat{u}}_b(t),~\vect{\hat{r}}_b(t)$ and $\vect{\pi}_b(t)$ for $t=0$ and $\forall b\in\Set{B}$
	\WHILE{ true }
		\STATE $t\rightarrow t+1$
		\STATE Action selection: $a_b(t) = f\Big( \vect{\pi}_b(t-1) \Big)$,~(\ref{eqn:prob_to_action})
		\STATE Load advertising: $\hat{\rho}_b(t)$,~(\ref{eqn:load_estimation})
		\STATE UE association: $b(\vectx, t)$,~(\ref{eqn:ue_association})
		\STATE Calculations: $ \rho_b(t),~\Gamma_b\big(\vect{\rho}(t)\big),~u_b\big(\vect{\rho}(t)\big) $
		\STATE Update utility and regret estimations, and probability:
		\STATE \hspace{10pt} $\vect{\hat{u}}_b(t+1),~\vect{\hat{r}}_b(t+1),~\vect{\pi}_b(t+1)$,~(\ref{eqn:algoUpdates})
	\ENDWHILE
\end{algorithmic}
\end{algorithm}

\section{Simulation Results}\label{sec:results}

For our simulations, we consider a single cell covered by a macrocell BS with an arbitrary number of SBSs and UEs uniformly distributed over the area.
All the BSs share the entire spectrum and thus, suffer co-channel interference.
We conduct a series of simulations for various practical configurations and the presented results are averaged over a large number of runs.
The parameters used for the simulations are summarized in Table~\ref{tab:sim_para}.
Moreover, we compare our proposed approach with the conventional network operation referred to hereinafter as ``classical approach'' in which BSs do not have the capability to switch between sleep-wake states.
For additional comparisons, the achievable lower bounds (the optimal cost-minimizing solution) are calculated using exhaustive search.

\begin{table}[!t]
\caption{Simulation parameters.}
\centering
\begin{tabular}{l c}
\hline
{\bf Parameter} & {\bf Value} \\
\hline \hline
Carrier frequency	& $2$ GHz \\
System bandwidth	& $10$ MHz \\
Thermal noise ($N_0$) &$-174$ dBm/Hz \\
Mean offered traffic $\big(\lambda(\vectx)/\mu(\vectx)\big)$ & 180 kbps\\
Maximum transmission powers: MBS, SBS & $46,~30$ dBm \\
\hline \multicolumn{2}{c}{\bf Minimum  distances} \\ \hline
MBS -- SBS, MBS -- UE & 75 m, 35m \\
SBS -- SBS, SBS -- UE & 40 m, 10 m \\
\hline \multicolumn{2}{c}{{\bf Path loss models} ($d$ in km)} \\ \hline
MBS - UE & $128.1 + 37.6\log_{10}(d)$\\
SBS - UE & $140.7 + 37.6\log_{10}(d)$\\
\hline \multicolumn{2}{c}{\bf Learning} \\ \hline
Boltzmann temperature ($\kappa$) & 10\\
Energy and load impacts on cost ($\alpha,~\beta$) & 0.5,~0.5 \\
learning rate exponents for $\tau, \iota~\mbox{and}~\varepsilon$ & $0.6,~0.7,~0.8$ \\
\hline
\end{tabular}
\label{tab:sim_para}
\end{table}

\begin{figure}[t]
\centering
\includegraphics[width=.44\textwidth]{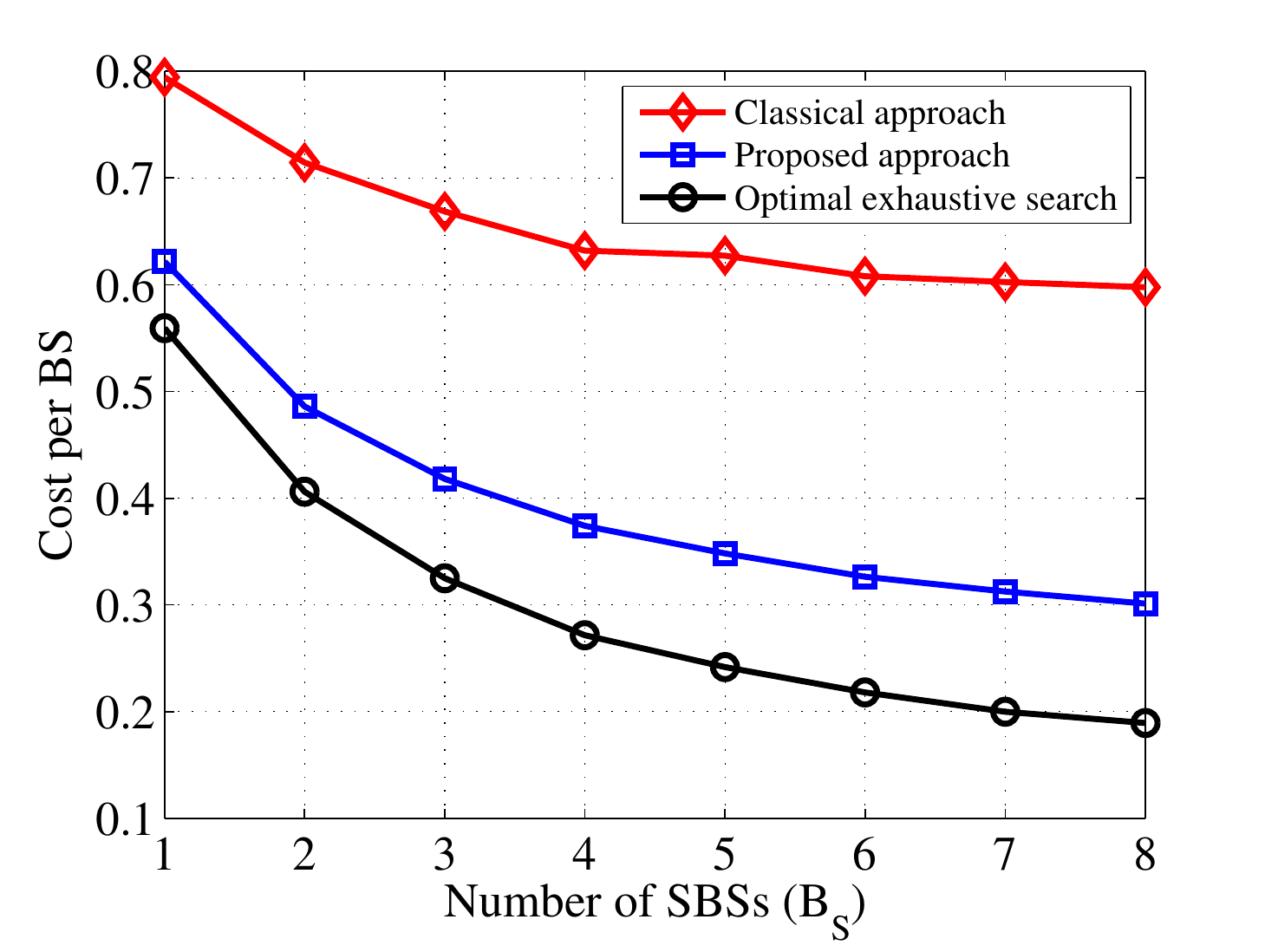}
\caption{Variation of the cost per BS with respect to the number of SBSs. The number of UEs is fixed to $100$.}
\label{fig:cost_vs_pbs_FB}
\end{figure}

Fig.~\ref{fig:cost_vs_pbs_FB} shows the average cost achieved per BS as the number of SBSs varies. The cost captures the tradeoff between load and energy consumption.
As the number of BSs increases, the total energy consumption of the network increases.
However, the load from a fixed number of UEs is distributed among the various BSs.
Therefore, the energy required to handle the load decreases per BS resulting in a cost decrease per BS for all three approaches  as seen in Fig.~\ref{fig:cost_vs_pbs_FB}.
In the proposed approach, the BSs switch to a sleep state when there are no UEs in their vicinity.
Consequently, Fig.~\ref{fig:cost_vs_pbs_FB} shows that the proposed approach exhibits a considerable cost reduction compared to the classical model.
IWe also see that, for a single SBS, the proposed approach exhibits a cost reduction of $21.8\%$ and it reaches up to $49.5\%$, relative to the classical approach, with eight SBSs.
Fig.~\ref{fig:cost_vs_pbs_FB} also shows that the difference in the average performance between the proposed approach and the optimal exhaustive search solution reaches about $18.8\%$ at $B_S =8$ SBSs.
However, the optimal solution requires an exhaustive centralized search which yields significant overhead. 
Indeed, the gap between the exhaustive search and the proposed approach is a byproduct of the uncoordinated decision making processes and the selfish behavior of the players (BSs).

\begin{figure}[t]
\centering
\includegraphics[width=.44\textwidth]{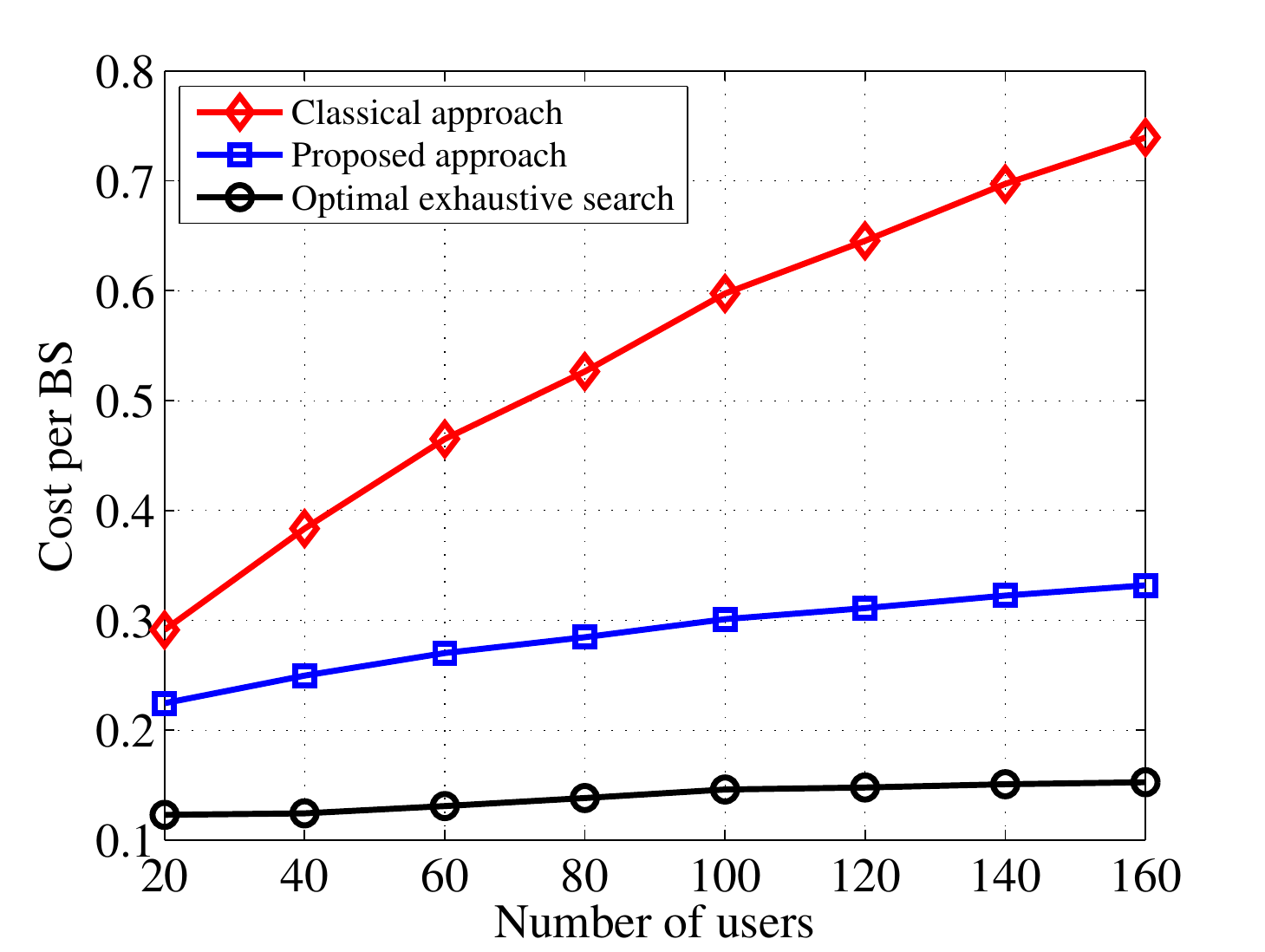}
\caption{Variation of the cost per BS as a function of the number of UEs. The number of BS is fixed to $8$.}
\label{fig:cost_vs_ue_FB}
\end{figure}

Fig. \ref{fig:cost_vs_ue_FB} shows the changes in the average cost per BS as the number of UEs varies.
With the increasing number of UEs, the load in the system increases and the BSs consume more energy to serve their UEs.
However, as seen in Fig~\ref{fig:cost_vs_ue_FB}, the proposed approach manages to reduce the cost by balancing the energy consumption and the handled load.
Fig. 3 demonstrates that the proposed learning approach yields significant cost reductions of up to $55\%$ (for a network with 160 UEs), relative to the classical approach.

\begin{figure}[t]
\centering
\includegraphics[width=.44\textwidth]{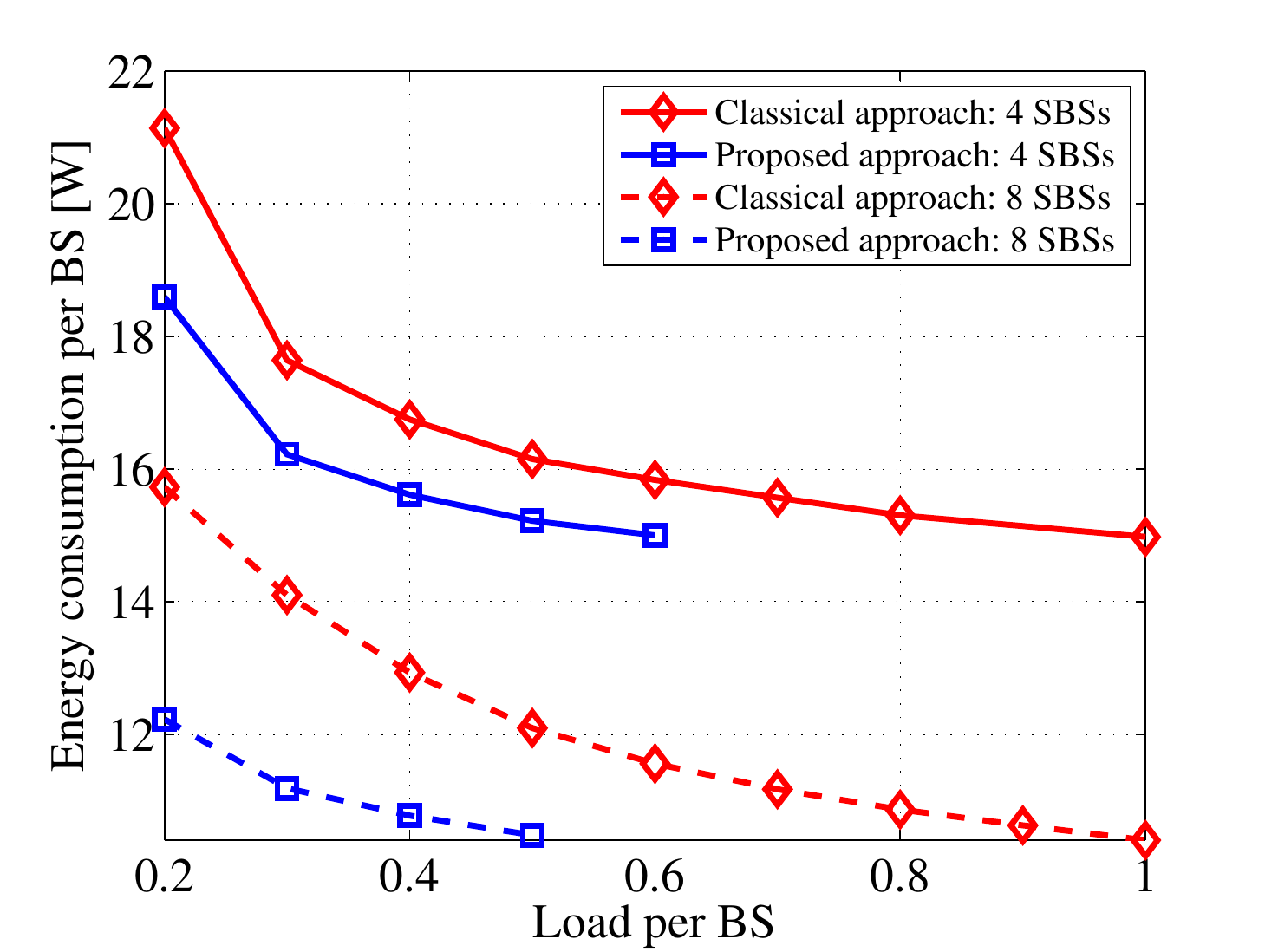}
\caption{The tradeoff between the load per BS and the energy consumption per BS for networks with 4 SBSs and 8 SBSs with various number of UEs.}
\label{fig:tradeoff}
\end{figure}

In Fig.~\ref{fig:tradeoff}, we show the tradeoff between energy consumption and load for both proposed and classical approaches for networks with 4 and 8 SBSs with various number of UEs.
Fig.~\ref{fig:tradeoff} corroborates the intuition derived in (\ref{eqn:state_based_power})-(\ref{eqn:load_bs}) which showed that, at the expense of increasing delays, energy savings are possible at higher loads.
Moreover, we can see that, for a given load, it can be observed that the proposed approach consumes less energy compared to the classical approach.
The main reason is that the opportunistic ON/OFF algorithm switches off unnecessary BSs and thus saves a large portion of energy compared to the classical approach. 
As we increase the number of SBSs from four to eight, the UEs are distributed among the additional BSs, and thus, both energy consumption per BS and load per BS are reduced.
Finally, Fig.~\ref{fig:tradeoff} shows that these energy reductions reach up to $10.8\%$ for four SBSs and $23\%$ for 8 SBSs compared to the classical approach.

\begin{figure}[t]
\centering
\includegraphics[width=.44\textwidth]{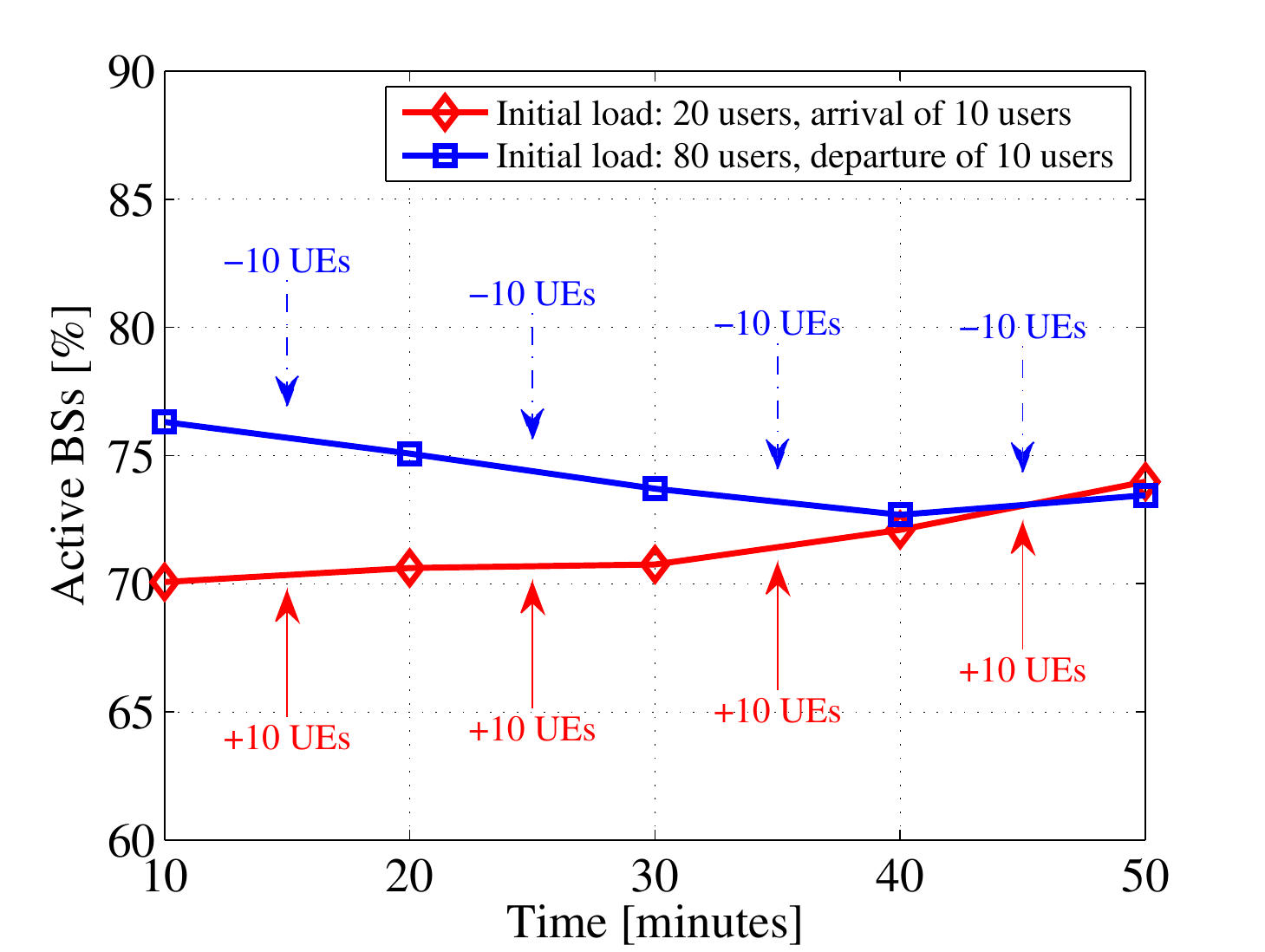}
\caption{Average percentage of active BSs as the network changes over time for two different cases of UEs departures/arrivals.}
\label{fig:dynamic_plot}
\end{figure}

Fig.~\ref{fig:dynamic_plot} shows how the proposed approach adapts to slowly varying network dynamics.
Here, the network starts with a given number of UEs.
After each 10 minutes, we evaluate the percentage of active BSs as the number of UEs changes. 
Here, we assume that each 10 minutes a maximum of 10 UEs will leave/join the network.
We consider two cases with five changes as follows:
{\it Case 1)} the simulation starts with 20 UEs.
Every 10 minutes, an additional 10 UEs will enter the network.
{\it Case 2)} the initial number of UEs is 80, and 10 UEs leave the network every 10 minutes. 
The results in Fig.~\ref{fig:dynamic_plot} show that the proposed approach can activate additional BSs to handle the increased load (case 1) as well as deactivate the unwanted BSs to save energy (case 2) thus adapting to the changes in the network composition.

\begin{figure}[t]
\centering
\includegraphics[width=.44\textwidth]{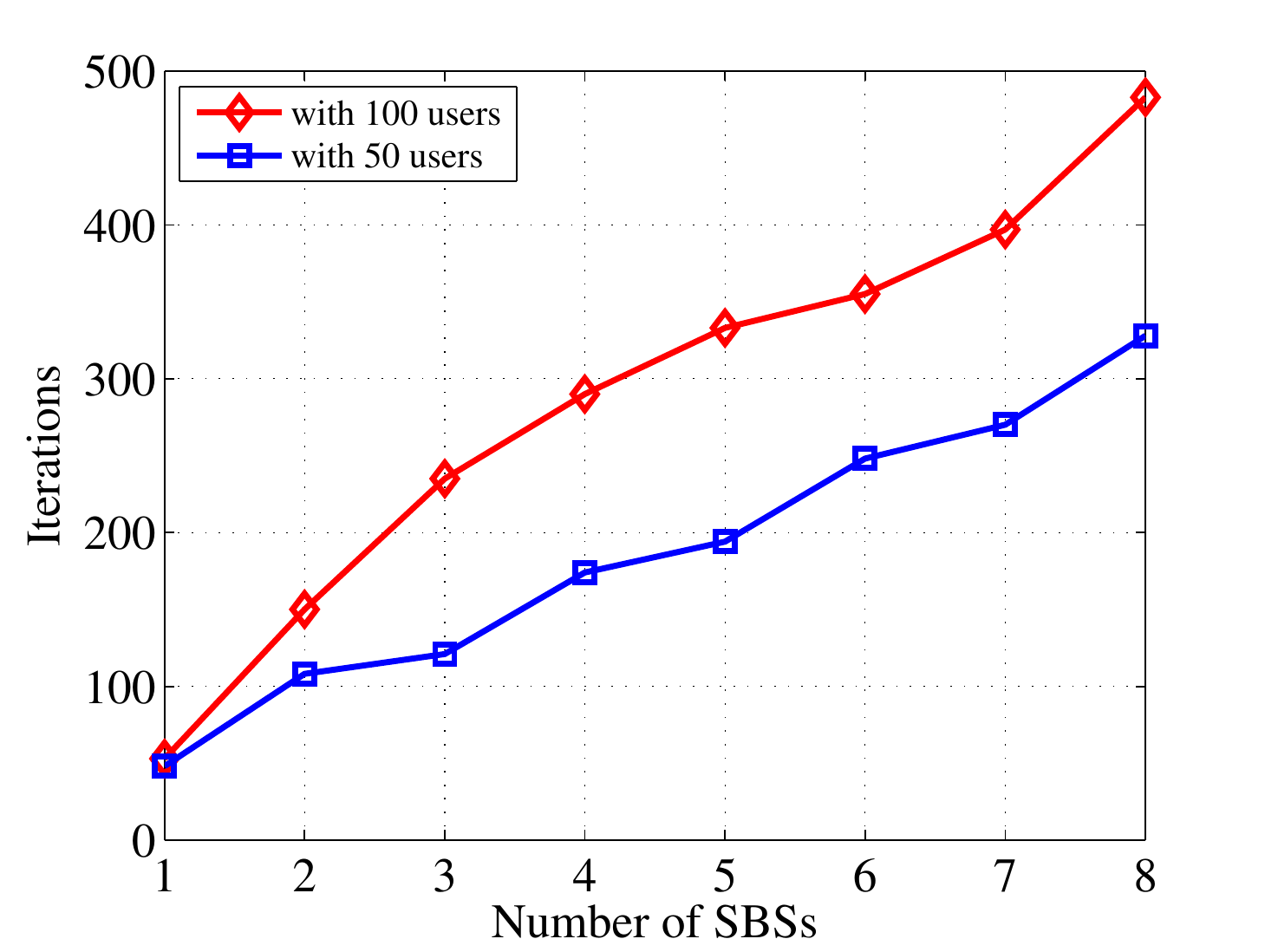}
\caption{The convergence time of the proposed algorithm as a function of the number of SBSs.}
\label{fig:converge_time}
\end{figure}

In Fig.~\ref{fig:converge_time}, we show the convergence time of the proposed approach as the number of SBSs varies for 50 and 100 users.
In this figure, we can see that, as the network size increases, the average number of iterations till convergence increases.
Fig.~\ref{fig:converge_time} also shows that reducing the number of UEs leads to a faster convergence time.
Although UEs are not players in the game, they affect load balancing among BSs for each action selection.
As the number of UEs increases, the frequency of offloading UEs among BSs increases, and, thus, a longer convergence time is observed. 
Fig.~\ref{fig:converge_time} shows that the maximum average number of iterations reaches up to $483$ for a network with $8$ SBSs and $100$ UEs.
\section{Conclusions}\label{sec:conclusions}

In this paper, we have proposed a distributed learning mechanism using which small cell base stations opportunistically switch between active and sleep modes, depending on various network parameters. 
We have formulated the problem as a game in which the goal of each base station is to minimize the system cost which captures the energy and load expenditures.
To solve the game, we have proposed a distributed algorithm using which the base stations choose their transmission modes with little additional overhead. 
Simulation results shows that the proposed learning approach yields significant gains, in terms of reducing energy consumption and overall cost in the network, when compared with conventional transmission techniques.
 The results also show that the proposed approach can successfully adapt to slow, periodic, and dynamic changes in the environment.

\bibliographystyle{IEEEtran}
\bibliography{IEEEabrv}

\end{document}